\documentstyle[prb,aps]{revtex}

\begin{document}
\draft

\twocolumn[\hsize\textwidth\columnwidth\hsize\csname @twocolumnfalse\endcsname

\title
{
The Equation of State and the Hugoniot of Laser Shock-Compressed Deuterium.
}
\author
{
 M.W.C. Dharma-wardana\cite{byline1}  and Fran\c{c}ois Perrot\cite{byline2}
}
\address
{
National Research Council, Ottawa,Canada. K1A 0R6
}
\date{-- Dec 2001}
\maketitle
\begin{abstract}
The equation of state and the shock Hugoniot of deuterium are
calculated using a  first-principles approach,
for the conditions of the
recent shock experiments.
We  use density-functional theory within a
a classical mapping of the quantum fluids
$\lbrack$Phys. Rev. Letters, {\bf 84}, 959 (2000)$\rbrack$.
 The calculated
Hugoniot is close to the Path-Integral Monte Carlo(PIMC)
result. We also consider the {\it quasi-equilibrium} two-temperature
case where the Deuterons are hotter than
the electrons; the resulting quasi-equilibrium
Hugoniot mimics the laser-shock
data. The increased compressibility
arises from hot $D^+-e$
pairs occurring close to
the zero of the electron chemical potential.
\end{abstract}
%\vspace{0.5in}
%\hspace{0.5in} see file /usr/people/chandre/tekstuff/hud/ms1.tex 
%
%\hspace{0.5in} date: \today 3-12-2001
\vskip2pc]
\narrowtext

  Hydrogen and its isotopes have been  extensively  studied, yet
 laser-shock experiments of Da Silva et al.\cite{dasilva},
Collins et al.\cite{collins}, and  Mostovych et al.\cite{mosto} 
produced unexpected disagreement
with the equation of state (EOS) of the
SESAME database.\cite{sesame}. 
The disagreement with SESAME occurs
 for  temperatures T
 with $\sim$ 0.8 eV $< T <$
 $\sim$ 10 eV, and
for densities  $1.8 < r_s < \sim 2$, where
the electron-density parameter
 (in atomic units) is,
 $r_s$ = $(3/4\pi\bar{n})^{1/3}$.
Here $\bar{n}$ is the electron-number density (au.).
 The coupling constant
 $\Gamma$=(potential energy)/(kinetic energy) ranges
from 1 to $\sim 30$.

The electrons change from a degenerate liquid to
a classical system in the
anomaly-regime (AR),
 while the $e-
D^+$ interaction is at the threshold of bound-state
 formation.
Thus the AR poses a  difficult, strongly-correlated
 many-body problem of wide interest -
 from astrophysics and fusion to materials science.
 Hence a flurry  of activity
has focused on the deuterium
EOS and its Hugoniot.\cite{hug}
These involve intuitive approaches 
(called ``chemical models''),\cite{ross}
assuming the existence of molecules etc.,
and first-principles approaches, e.g., density-functional theory (DFT),
 or  quantum Monte Carlo (QMC). 
  Standard DFT 
 treats the ions as
an external potential,  optimized via
 Car-Parrinello type
techniques, while the
 local-density-approximation
(LDA), or  simpler
 tight-binding approaches are used for the electrons.\cite{dftcalcs,lenox}
Path-integral
Monte Carlo (PIMC),
a  finite-T QMC approach, has been used for this problem.\cite{pimc} 

PIMC and DFT calculations 
yield shock Hugoniots close to SESAME,
  showing no 
 strong softening.
 Recent experiments from Sandia
are also close to the SESAME Hugoniot.\cite{knud}
 However  a high degree of agreement exists
in the laser-shock results.\cite{dbleshock}

Our objective is to present new
 first-principles results
for the EOS and Hugoniot of deuterium, using methods 
quite independent of
previous methods. Our method 
is computationally simple and uses
one-dimensional integral equations
based on DFT. We  calculate
all the pair-distribution functions (PDFs),
 $g_{ij}$, with
$i=$1,2,3; (or e$_\uparrow$, e$_\downarrow$, and D$^+$ nuclei)
 of a three component
 system.
 Being based on the PDFs, it is manifestly  non-local
 and has no self-interaction errors.
 It  easily adapts to the
quasi-stationary two-temperature case with
$T_D\ne T_e$.
Consider the fluid with a deuterium nucleus at the origin,
and let the one-body  densities  of the electrons and $D^+$ be
  $n(r)$ and $\rho(r)$.
Then  $n(r)$
is really $n_{De}(r)=\bar{n}g_{De}(r)$. The free energy $F$
 is a functional of the form
 $F[n(r),\rho(r)]$. Thus, taking functional
derivatives,  we have {\it two} coupled Kohn-Sham-Mermin equations:
\begin{eqnarray}
\delta F[n(r),\rho(r)]/\delta n(r)&=&0 \\
\delta F[n(r),\rho(r)]/\delta \rho(r)&=&0
\end{eqnarray}
 As shown in
ref.~[\onlinecite{hyd0,ilciacco}]
Eq.(1) leads to
a (quantum) Kohn-Sham (K-S) equation for the electrons,
 while Eq.(2),
 a classical K-S
equation, becomes the Hyper-Netted-Chain (HNC)\cite{hncref}
 equation for a specific 
choice of the correlation potential (there being no exchange potential
in the classical system). Thus the K-S eigenfunctions, as well as the
$n(r)$, $\rho(r)$, in the Hydrogen fluid  were calculated by
 solving Eqs. (1-2).\cite{hyd0}
In Car-Parrinello approaches only
 the electrons are treated in
DFT, while the N ions at sites $\vec{R_i}$ are explicitly treated.
Here  both electrons and ions are treated using
 their distributions.
 Our hydrogen
calculations were later
confirmed by lengthy QMC.\cite{kwon}
However, while Eqs.(1-2) provide the $g_{ie}(r)$, $g_{ii}(r)$, i=ion,
 and the K-S
states in the fluid,
the $g_{ee}(r)$ was available only in LDA.

 Recently, we showed how the electrons at the
 physical temperature $T_e$
could be replaced by an equivalent classical system at
$T_{ee}=(T_e^2+T_{eq}^2)^{1/2}$,
 such that
 the quantum effects
are correctly incorporated.
A simple expression for the electron ``quantum temperature'' $T_{eq}$
 as a function of $r_s$  was given
in ref.\onlinecite{prl}.
 Application of the method, denoted the ``Classical mapping
 of quantum systems
using the  Hyper-Netted-Chain equation (CHNC)'', to
3-D and 2-D uniform electron liquids at
$T=0$ and finite-$T$,
showed  excellent agreement
 of the $g_{ij}$, 
energies, etc., with QMC results, for even very strongly
coupled situations.\cite{prl,prb00,2D} 
Deuterons in a uniform neutralizing background
are mathematically identical to the electron system,
 except for changes of scale,
(e.g, for deuterons, $r_{sD}=r_s(M_D/m_e)$ etc.,
$M_D$ is the deuteron mass and $m_e=1$).
Hence the $D^+$-quantum temperature $T_{Dq}$ is also available, and
 is negligible in the
 regime of interest; thus $D^+$ are treated as classical particles.

 The mean
densities $\bar{\rho}$ and $\bar{n} $ are equal since
 the nuclear charge $Z$ = 1.
Consider a fluid of total  density $n_{tot}$, with
three species. Let
 $x_i$ = $n_i/n_{tot}$, $n_{tot}=\bar{\rho}+\bar{n}$. 
The physical temperature is  $T$,
while the inverse temperature
of the electrons
is $1/\beta_{ee}$,
with  $1/\beta_{ee}$= ${\surd{(T^2+T_q^2)}}$.
For $D^+$, the classical mapping $T_{DD}=1/\beta_{DD}$ is
taken to be  $T$.
The inverse temperature
  $\beta_{ij}$ between electrons
and deuterons is actually  not needed since the {\it product},
 $\beta_{eD}\phi_{eD}(r)$
 is completely determined
 by the quantum mechanics
of the problem (see below).
 However, for the present application
it is the mean kinetic energy of
the $D^+-e$ pair, i.e., $T_{eD}=(T_{ee}+T_{DD})/2$, consistent with the
classical picture used here.

The classical equations for the PDFs
and the Ornstein-Zernike(OZ) relations are:
\begin{eqnarray}
\label{hnc}
g_{ij}(r)&=&exp[-\beta_{ij}\phi_{ij}(r)
+h_{ij}(r)-c_{ij}(r) + B_{ij}(r)]\\
 h_{ij}(r) &=& c_{ij}(r)+
\Sigma_s n_s\int d{\bf r}'h_{i,s}
(|{\bf r}-{\bf r}'|)c_{s,j}({\bf r}')
\label{oz}
\end{eqnarray}
Here $\phi_{ij}(r)$ is the pair potential between the
species $i,j$. For e-e (or $D^+-D^+$) this is
just the Coulomb potential $V_{cou}(r)$.
If the electron spins are parallel, the Pauli
principle prevents  occupation of the same spatial orbital.
As before,\cite{prl}
 we introduce a
``Pauli exclusion potential'', ${\cal P}(r)$.
Thus $\phi_{ij}(r)$ becomes ${\cal P}(r)\delta_{ij}+V_{cou}(r)$,
when $i,j$ denote electrons.
The function $h(r) = g(r)-1$ is related to the
structure factor $S(k)$ by a Fourier transform.
The  $c(r)$ is the ``direct correlation function (DCF)''
of the OZ equations.
The $B_{ij}(r)$  is
the ``bridge'' term due to certain cluster interactions.
If this is neglected,
Eqs.~\ref{hnc}-~\ref{oz}  form a closed set defining the
HNC approximation. (In effect, the classical K-S equation
becomes the HNC equation if the correlation potential
is evaluated as a sum of hyper-netted-chain diagrams.)  
The HNC is sufficient for the uniform 3DEG for a
range of $r_s$, up to $r_s=50$, as studied previously.\cite{prb00}
We neglect the bridge corrections in
this study of deuterium.

 The ${\cal P}(r)$ is defined as in ref.\onlinecite{prl} from the
zeroth-order PDFs of the parallel-spin electrons. Thus:
\begin{equation}
\beta{\cal{P}}(r)=h_{11}^0(r)-c_{11}^0(r)-ln[g_{11}^0(r)]
\end{equation}
where, e.g., $c^0_{11}(r)$ is the spin-$\|$ DCF
of the O-Z equation.
Only the
product $\beta{\cal{P}}(r)$ is needed.

The Coulomb potential $V_{cou}(r)$
for two point-charge electrons is  $1/r$.
However,  an electron at
the temperature $T$ is 
localized to within a thermal  de Broglie length (dBL). Thus,  
for the 3DEG we used a ``diffraction
corrected'' form:\cite{minoo},
%\begin{eqnarray}
\begin{equation}
\label{potd}
V^{ee}_{cou}(r)=(1/r)[1-e^{-rk_{ee}}]
\end{equation}
where 
$ k_{ee}=(2\pi m^*T_{ee})^{1/2}$ as shown by Minoo et al.\cite{minoo}
Here $m^*$=1/2 is the reduced mass of the electron {\it pair}.
In Minoo et al, the physical temperature $T$ was 
used and is
valid  only at high $T$.
 The use of $T_{ee}=(T^2+T_{eq}^2)^{1/2}$
instead of $T$ validates it down to $T=0$ as well.\cite{prl}

Since the $D^+$ are classical particles,
the $D^+-D^+$ interaction is  the 
coulomb interaction $\phi_{33}(r)=1/r$.
The $e-D^+$ interaction is more tricky.
 The
e-e interaction, eqn.~\ref{potd}, is based on the
 quantum mechanical scattering of two electrons.
  We determine the  $e-D^+$ interaction
 $\phi_{eD}(r)$, from the density profile $n_{De}(r)$ 
given by the K-S equation for 
electrons interacting with {\it  a single deuteron} at the origin.
We have discussed this in the
context of the ``neutral pseudo-atom'' DFT model (NPA-DFT) for solving
the K-S equations.\cite{ilciacco,perrot93}
 This gives 
the deuteron-electron PDF, i.e., 
$g_{De}(r)$ = $n_{De}(r)/\bar{n}$. This 
 $n_{De}(r)$ includes {\it both bound-state and continuum-state}
densities.
Applying the HNC 
and the OZ equation to this system containing a {\it single}
 deuteron, we have,
\begin{eqnarray}
\label{potDe}
-\beta_{De}\phi_{De}(r)&=& log[g_{De}(r)]-h_{De}(r)+c_{De}(r)\\
h_{De}(r)&=&c_{De}(r)+ \bar{n}\int d\vec{r'}c_{De}(\vec{r}-\vec{r'})h _{ee}(r')
\end{eqnarray}
The deuteron-deuteron DCF does not appear
as there is only a single $D^+$. Hence, knowing the $g_{De}(r)$
 from the solution of the
Kohn-Sham equation for the single deuteron problem, we can obtain $c_{De}(r)$ in
terms of $h_{De}(r)$. Hence the e-D potential can be extracted. 
This determines the product,
 $\beta_{De}\phi_{De}(r)$, while $\beta_{De}$, and  $\phi_{De}$ are not needed
individually.
However, on solving the K-S equation for the regime of interest,
{\it no atomic  bound
states} are found; the effective ionic charge $Z-n_b$,
 where $n_b$ is the number of bound electrons
per nucleus, remains unity.
This does not  contradict 
{\it transient} bound-states found in PIMC.\cite{lqdmets}
 Hence, at least in this regime of $\bar{n}$  and $T$,
Kohn-Sham NPA-DFT is not needed;
 We set:
\begin{eqnarray}
\phi_{De}(r)&=&-(1/r)[1-e^{-rk_{De}}]\\
k_{De}&=&(2\pi m_{e}T_{ee})^{1/2}\\
1/\beta_{De}&=&(T_{DD}+T_{ee})/2
\end{eqnarray}
The first equation is just the $V_{cou}$ with the $r=0$
 value set to an
inverse dBL for the D-e pair, $k_{De}$,  as
in the e-e interaction.
  The dBL $1/k_{De}$
contains only the electron contribution
since the D contribution is zero
for a classical
particle. Thus only the $T_{ee}$ appears in $k_{De}$
(the effective
mass of the D-e pair is  $m_e$,  since the deuteron mass $M_D >> m_e$).
The effective temperature $1/\beta_{De}$ for the D-e interaction is
  the mean kinetic
energy of the pair.
However, the use of the K-S procedure for the
dimensionless potential  $\beta_{De}\phi_{De}(r)$
becomes obligatory at lower densities ($r_s > 2$) and
 temperatures ($T < 1.0$ eV) when
the bound-state population $n_b$ becomes nonzero.

	Using the above scheme, we solve the coupled set
 of HNC equations to determine
the six PDFs of
the $e,D^+$ system.
 The excess free-energy
 $F_{exc}(r_s,T)$
is determined via a coupling-constant integration,
 as in ref.~[\onlinecite{prb00}],
 for a range of values of $r_s$
 and $T$, and converted into the total free energy $F(r_s,T)$ by
 adding on the
ideal electron and ion contributions $F_0^D,F_0^e$ (see Fig.1(a)).
 The total pressure $P$ and the
total internal energy $E$
 are obtained as usual by
$P=\partial F(r_s,T)/\partial V$, where $V$ is the volume, and
 $E=\partial \beta F(r_s,T)/\partial \beta$, where $\beta=1/T$.
In the regime of interest, i.e, 1.8 $<r_s<$ 2.1, and 
0.8 eV $< T <$ 15,
 we find that $F_{exc}(r_s,T)$ is
approximately linear, i.e., $F_{exc}(r_s,T)=M(T)r_s+C(T)$.
The $T$-dependence of $M(T)$, $C(T)$ is quite nonlinear. 
Figure 1(b) shows that $M(T)$  changes character near
$T$ = 4 eV.

Our $P$, $E$ results are compared with the PIMC
of Militzer and Ceperley (MC) in table I,
showing good agreement for $T > $ 2.75 eV. For lower $T$,
our pressures are smaller.

	The free-energy  $F(r_s,T)$ is used to calculate 
the deuterium Hugoniot for the
initial state, ($E_0,V_0,P_0$), with $T$ = 19.6 K and an
 initial density $\rho_0$ = 0.171 g/cm$^3$.
 The  initial-state $E_0$ = -15.886 eV per atom, and $P_0=0$.
The CHNC result, and others are
shown in Fig.2.
The CHNC Hugoniot, similar to PIMC,
approaches SESAME
at high temperatures. A softening of the 
Hugoniot around 2 Mbar, not seen in the PIMC curve is also noted.
 This
appears near $\bar{n}$ and $T$ 
 where the interacting electron chemical potential
$\mu_e(r_s,T)$ passes through zero.

	We  turn to the HNC equations to consider
electrons and deuterons at two different temperatures,
 $T_e$ and $T_D$. The
shock is launched from an Aluminum pusher; the ions are
initially much hotter than the electrons.
 The
velocity  measurements begin after about 3 ns
 in the laser experiments, and after a longer time
in the Sandia work. 
 Landau-Spitzer theory would indicate that the
 $D^+-e$ equilibriation
occurs well within nanosecond timescales.
 However, the formation
 of coupled modes in plasmas with $\Gamma > 1$
strongly quenches the ion-electron equilibriation process.\cite{elr}
Also, experimental evidence exists for this point of view.\cite{elrexp}
A compactly held screening-charge  at each ion would
act like a neutral object which, while having a very hot
deuteron at the centre, would screen it
from the cooler outer electrons.
 The effect could lower the
electron-ion relaxation by more than an order of magnitude.\cite{elr,elrexp}
In lieu of a systematic energy relaxation analysis,
here we
assume  that the deuteron nuclei are about 5 eV hotter
than the nominal electron temperature ( i.e., $T_D = T_e+5$ eV),
 except at the
lowest temperatures.
Using $T_e$, $T_D$ in the HNC equations as before,
we have calculated
a quasi-equilibrium $F_{exc}(r_s,T_e,T_D)$
 and a shock Hugoniot (quasi-equilibrium concepts
are discussed
in ref.\onlinecite{elr}).
The resulting non-equilibrium
Hugoniot is given in Fig.2.

The higher $T_D$ makes
 the deuteron-electron fluid
more compressible. This appears counter-intuitive if one 
considers only the $D^+$ contribution.
 The quasi free energy $F(r_s,T_e,T_D)$ consists of
 $F_e$, $F_D$, and $F_{De}$. On setting $T_D > T_e$, the 
$F_D$ term taken alone {\it reduces} the compressibility, but the total
 compressibility
is increased by the major role of the pair-term $F_{De}$.
 As seen in Fig.1~(a), the
fluid is in a regime  close to the
 $\mu_e= 0 $ transition.
 Thus a higher $T_D$ increases $T_{De}$ and
 reduces the electron degeneracy even more, 
making it more compressible. This also reduces the screening
and makes the electrons interact more strongly with the nuclei.
When this effect is strong
enough to offset the reduction of the compressibility from the ideal
gas term  of the hotter deuterons, a softening of the Hugoniot
could result.  In Fig.2 we  show a Hugoniot labeled
NEQ0 where  the ideal term $P_0$ was computed just
as in the equilibrium Hugoniot, while in NEQ the full effect was included.
Thus we see that except for the lowest temperatures, the contribution of
the deuteron-electron pairs dominate.

Our explanation of the observed
laser-shock Hugoniot is incomplete until
detailed modeling of the passage to equilibrium
is achieved. Other factors like the planarity, constancy and
duration of the shockwave are also relevant.
 However, the present
 discussion strongly justifies the need
to include equilibriation effects as well.
 The CHNC approach presented here is numerically and
computationally very simple and should be a handy tool
in such studies. In fact all the calculations presented
here were carried out using very modest computational
facilities.\cite{sgi}
Our computer codes  may be remotely
accessed by interested researchers by visiting our website.\cite{web}

	In conclusion, we present a parameter-free calculation
of the EOS of deuterium in the regime of densities and
temperatures addressed by recent laser and magnetic
shock experiments. Kohn-Sham calculations show the absence of
atomic  bound states in this regime.
Hence a  simple 
classical mapping of the electron quantum fluid was used.
 The calculated Hugoniot
is in good agreement with other first-principles calculations.
Calculations for the case with the ions hotter
than the electrons by about 5 eV are also presented. They
 suggest the possibility
 that the anomalous Hugoniots obtained
from laser experiments could arise from the slowness of the
system to equilibriate within the experimental time scales.

%\newpage    %%%%%%%%%%%%%%%%%%%%%%%%%%%%%%%%%%%%
%\twocolumn  %%%%%%%%%%%%%%%%%%%%%%%%%%%%%%%%%%%%
%\narrowtext %%%%%%%%%%%%%%%%%%%%%%%%%%%%%%%%%%%%
%

\newpage

\begin{table}
\caption{
The total pressure $P$ (Mbar) and total energy $E$ (eV),
calculated using the Classical-map HNC (CHNC) approach, and the
path-integral Monte Carlo (PIMC)
approach of Militzer and Ceperley (MC) at $r_s$=2.0, i.e, at
a deuterium density of 0.6691 g/cm$^3$. 
}
\begin{tabular}{ccccccccc}
$T(K)$&$F_{exc}$&$P$(CHNC)&$P$(MC)&$E$(CHNC)&$E$(MC)\\
\hline\\
 500 000& -5.35310 & 26.278 & 25.980  &  113.30  & 113.20 \\
 250 000& -2.14960 & 12.244 & 12.120  &   47.57  &  45.70 \\
 125 000& -0.99712 &  5.374 &  5.290  &   13.60  &  11.50 \\
  65 000& -0.64405 &  2.143 &  2.280  &   -3.21  &  -3.80 \\
  31 250& -0.57058 &  0.754 &  1.110  &  -10.74  &  -9.90 \\
  15 625& -0.57119 &  0.213 &  0.540  &  -13.97  & -12.90 \\
  10 000& -0.57890 &  0.058 &  0.470  &  -14.91  & -13.60 \\
\end{tabular}
\label{gamma0}
\end{table}
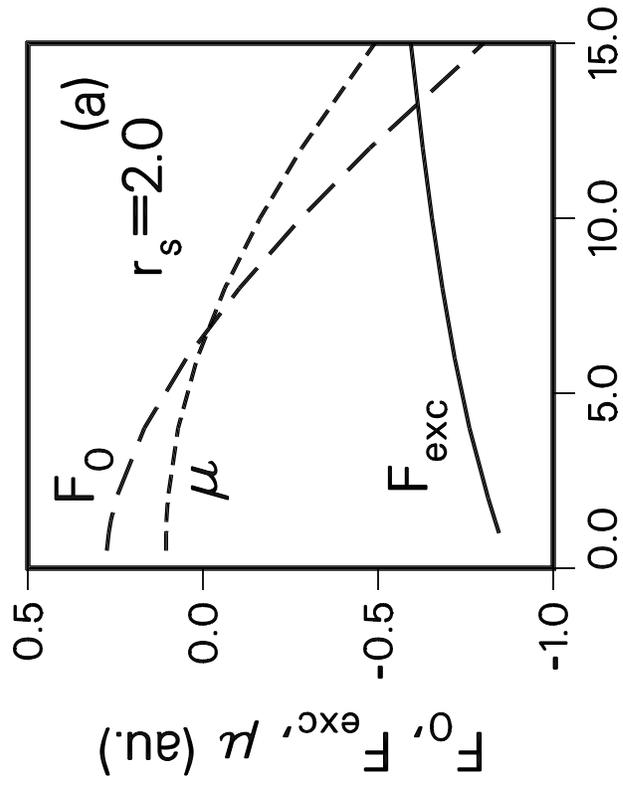
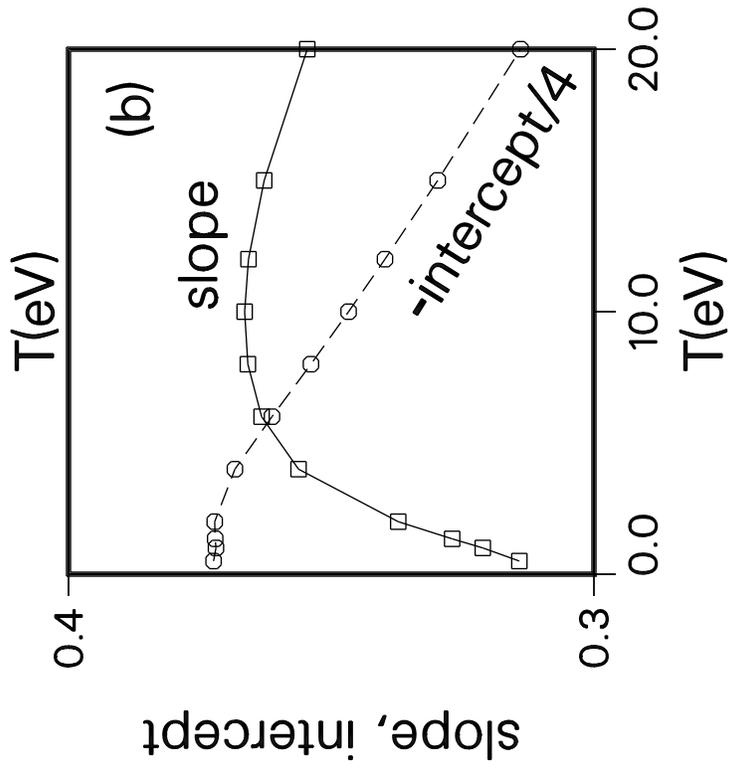
\begin{figure}
\caption
{(a) The temperature dependence of the excess free energy $F_{exc}(r_s,T)$,
$F_0^e((r_s,T)$, and the interacting chemical potential $\mu_e(r_s,T)$,
at $r_s$=1.86.  Note the passage of $\mu$
through zero.
(b) The excess free energy in the range of the the shock experiments
can be fitted to the linear form $F_{exc}(r_s,T)=M(T)r_s+C(T)$;
the slope $M(T)$ and the intercept $C(T)$ are shown as a function of
$T$. Nore the change of character in $M(T)$ when 
 $\mu_e$ passes through zero.
}
\label{figFexc}
\end{figure}

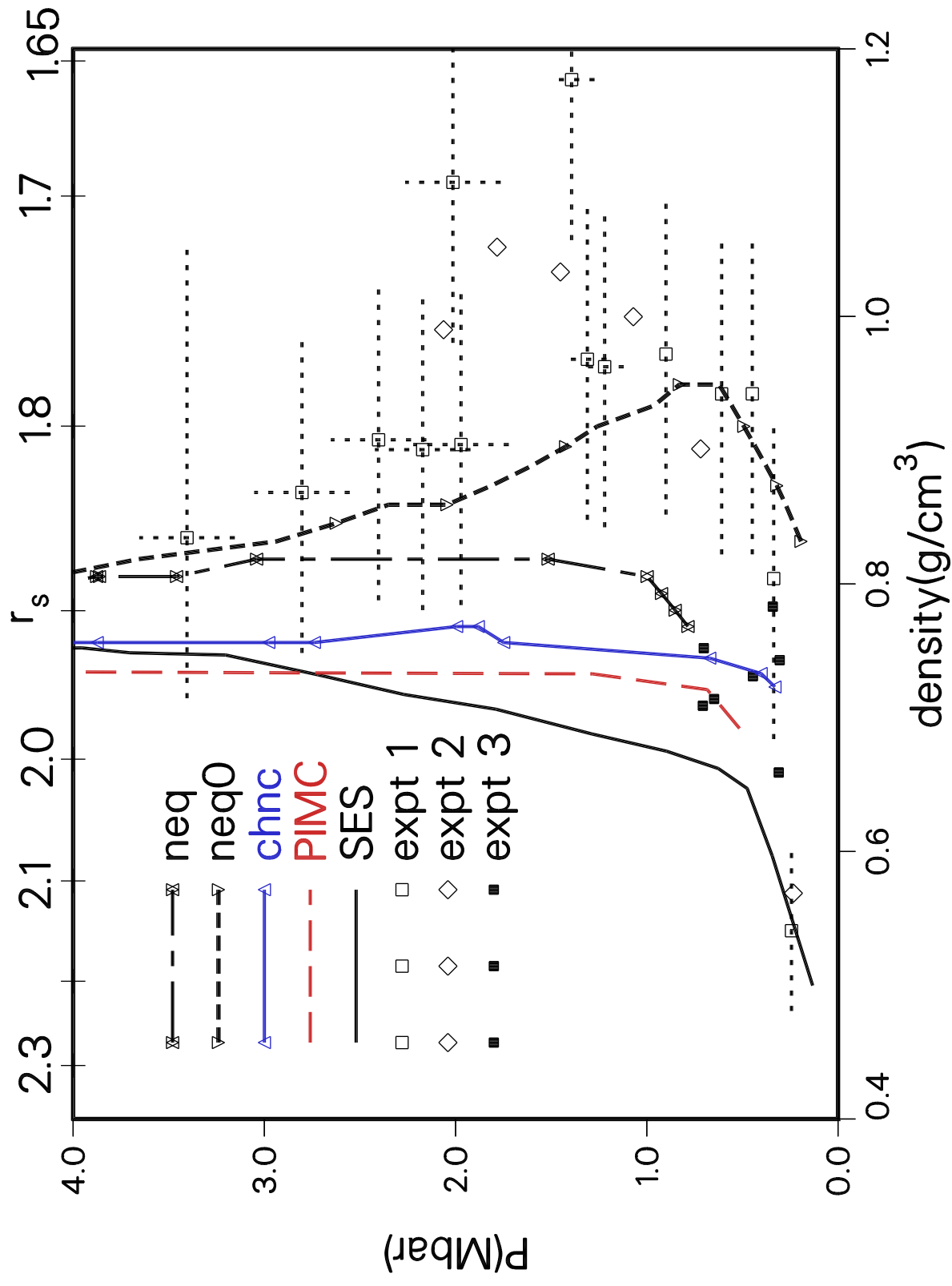
\begin{figure}
\caption
{ Comparison of the CHNC Hugoniot with experiment and other
  theories. Two non-equilibrium Huoniots are also shown (see text).
  Experiments 1, 2 and 3 refer to Da Silva {\it et al.}, 
Collins  {\it et al.}, and Knudson  {\it et al.}, respectively.
}
\label{fighu}
\end{figure}


\begin{references}
\bibitem{hug}
Hugoniot is the locus of states ($P,\rho,T$) generated by shock compression
from a given initial state.
\end{references}
\end{document}